\documentstyle[aps]{revtex}

\begin{document}

\title{Thermodynamic properties of the \(SO(5)\) theory for the 
antiferromagnetism and \(d\)-wave superconductivity:
a Monte Carlo study} 

\author{Xiao Hu}

\address{National Research Institute for Metals, 
Tsukuba 305-0047, Japan}

\date{submitted to Phys. Rev. B on \today}

\maketitle

\begin{abstract}
Thermodynamic properties of the \(SO(5)\) theory unifying the 
antiferromagnetism (AF) and the \(d\)-wave superconductivity (SC) are 
explored by means of Monte Carlo simulations on a classical model 
hamiltonian. The present approach takes into account thermal 
fluctuations both in the rotation of \(SO(5)\) superspins between the
AF and SC subspaces, and in the phase variables of SC order parameters.
Temperature vs. \(g\)-field phase diagrams for
null external magnetic field are presented, where the 
\(g\) field is conjugate with the quadratic order parameters and breaks
the \(SO(5)\) symmetry.  The normal(N)/AF and N/SC phase boundaries, 
both associated with second-order phase transitions, merge tangentially 
at the bicritical point into the
first-order AF/SC phase boundary. Hysteresis phenomenon is observed 
at the AF/SC phase transition, and therefore the present study suggests
the existence of a phase-separation region in the phase diagram.
Enhancement of AF correlations is observed above the SC critical 
temperature near the bicritical point in systems with AF couplings
stronger than SC ones.  Its relation with the spin-gap 
phenomenon is addressed. 
The \(SO(5)\) theory in an external magnetic field is also investigated,
and the following properties are clarified:  At sufficiently large \(g\) 
fields the SC order is established through a first-order freezing 
transition from the flux-line liquid into the flux-line lattice. 
Short-range AF fluctuations are larger at cores of flux lines than elsewhere, and decrease continuously to zero with increasing \(g\) field.
At intermediate \(g\) fields, the flux-line 
lattice of long-range SC order  and the long-range AF order coexist. 
Superlattice spots surrounding the strong AF Bragg peaks at
\(Q=(\pm\pi,\pm\pi)\) are observed in the simulated structure factor,
and are identified with the modulation by the triangular flux-line 
lattice of SC.  The AF phase boundary associated with the 
continuous onset of long-range AF order drops sharply to the \(g\) axis 
from a finite temperature in the temperature vs. \(g\)-field phase 
diagram.
  
\vskip1cm

\end{abstract}

\newpage

\section{Introduction}

High-\(T_c\) superconductivity (SC) in cuprates \cite{BM} is 
achieved by hole doping from the insulating state of
 antiferromagnetism (AF).  The AF and SC phases are proximate each 
other in temperature vs. hole-doping-rate phase diagrams.  
Enhancement of AF correlations is observed above the SC transition
temperature in the underdoped region \cite{Yasuoka}. 
A clear signal from these experimental facts is   
the importance of the inter-relationship between these two very
different and even expelling, at a first glance, properties.  
To explain theoretically the complex phase diagrams of the high-\(T_c\)
SC is still very challenging for the condensed matter physics.
This problem has been approached using microscopic models,
such as Hubbard hamiltonian and the simplified \(t-J\) hamiltonian
\cite{Anderson,ZR}. It has been tried to derive microscopically
the attractive force necessary for Cooper-pair formation from the magnetic interactions, and to construct the phase diagram with  the AF
and SC phases side by side.  Up to date, however, there is no well 
accepted 
microscopic theory which can count for the most important features of 
the high-\(T_c\) SC both qualitatively and quantitatively.

In the \(SO(5)\) theory this problem is approached in another way 
\cite{Zhang,Demler}:  The long-range SC and AF 
orders are presumed as the two possible long-range orders in pure 
systems.  The three components of the AF order parameter and 
the real and 
imaginary parts of the SC order parameter compose a five-component
superspin of \(SO(5)\) symmetry.  At low temperatures the  \(SO(5)\) symmetry is broken into two subspaces, the \(SO(3)\) one
associated with AF, and the \(U(1)\) one associated with SC.  The 
destination of the broken symmetry is controlled by the doping rate, 
or the chemical potential of holes.  
Therefore, the doping rate plays the role of 
\(SO(5)\)-symmetry breaking field.  Much interest has been 
stimulated by the proposal of the \(SO(5)\) theory, and considerable 
progresses in exploring this theory have been achieved 
\cite{Meixner,Arovas,Koyama,Rabello,Henley,Eder,Huso5,Nagaosa,Alama}. 
Since the superspin vector in the \(SO(5)\) theory is of five dimensions, 
three for AF and two for SC, entropy effects on the 
competition between these two long-range orders are highly nontrivial.  
Thermal fluctuations are very crucial in determining phase diagrams 
\cite{Huso5}. Therefore, investigation of 
the \(SO(5)\) theory at finite temperatures is essential 
for a comprehensive understanding of the theory.  Ultimately one should compare the predictions by the theory with phase diagrams observed experimentally. The \(SO(5)\) theory also raises interests in the point of 
view of phase transitions and critical phenomena.
To reveal the thermodynamic properties of the \(SO(5)\)
theory is the objective of the present study.    
Another important issue is the
competition between the long-range SC order in the presence of an
external magnetic field, realized in the flux-line lattice (FLL), and the
long-range AF order.
Since many interesting magnetic-field responses have been 
clarified in the long-range AF and SC orders separately, the proximity of 
them in high-\(T_c\) cuprates is very likely to produce more 
sophisticated phenomena.  Actually, it is suggested that
vortices induced by an external magnetic field in high-\(T_c\) 
superconductors may possess AF cores 
\cite{Zhang}.  To explore the vortex states in high-\(T_c\) cuprates
in the scheme of SO(5) theory is also very important.

In order to achieve the above purposes, Monte Carlo simulations on a 
classical model hamiltonian in three-dimensional (3D) space are 
performed \cite{Huso5}.  The present approach takes into account 
thermal fluctuations both in the rotation of \(SO(5)\) superspins 
between the AF and SC subspaces, and in the phase variables of SC order 
parameters.  The remainder of this 
paper is organized as follows: The hamiltonian is presented in Sec. II, 
with descriptions on technical details of simulation.
In Sec. III, simulation 
results for the null external magnetic field are presented.  There are
the AF, AF and SC phase-separation, and SC phases in the phase 
diagrams.  The spin-gap phenomenon is also addressed.
Section IV is devoted to reveal the effects of an external magnetic field 
in the \(SO(5)\) theory.  Coexistence between the long-range AF order
and the FLL of long-range SC order is observed. 
Vortex cores are found of larger AF components 
than elsewhere. Summary is given in Sec. V.

\section{Hamiltonian and simulation techniques}

The hamiltonian in the present study is given by

\begin{equation}
 {\cal  H}
  =-\sum_{\langle i,j\rangle}J^{SC}_{i,j}{\bf t}_i\cdot{\bf t}_j
    +\sum_{\langle i,j\rangle} J^{AF}_{i,j}{\bf s}_i\cdot{\bf s}_j
     +g\sum_i{\bf s}^2_i,
\label{eqn:H1}
\end{equation}

\noindent defined on the simple cubic lattice.
The vector {\bf t}, of two components and coupling ferromagnetically
with nearest neighbors, is for the \(d\)-wave SC order parameter; 
the vector {\bf s}, of three components and with AF coupling between
nearest neighbors, is for the AF order parameter.  The interplay
between the SC and AF order parameters is 
introduced by the \(SO(5)\) constraint on the superspin:

\begin{equation}
{\bf s}^2_i+{\bf t}^2_i=1.
\end{equation}

\noindent The \(g\) factor is a field  breaking the \(SO(5)\) symmetry 
into the \(U(1)\) and \(SO(3)\) subgroups, and is proportional to the 
doping rate in a loose sense \cite{Zhang}. 

The following notes on the above hamiltonian seem appropriate at this 
stage.  First, the above hamiltonian can be considered as the 
Ginzburg-Landau description of the \(SO(5)\) theory.
Both of the AF and SC order parameters, {\bf s} and {\bf t}, are
defined in a scale larger than the atomic one, but much smaller than
the macroscopic one.  In this sense they should be called as the local
order parameters.  The constraint (2) does not imply the existence
of long-range order in the macroscopic scale.  The long-range order 
parameter for the AF component is the staggered magnetization, and
that for the SC component is the helicity modulus \cite{Fisher2}. Second,
although no quantum effect is included explicitly in hamiltonian (1), 
the competition between the two different long-range orders
is taken into account sufficiently.  Therefore, the profound, 
nontrivial thermodynamic properties of the \(SO(5)\) theory can be captured. Third, thermal fluctuations in 
phase variables of SC order parameters, which are especially
important for underdoped high-\(T_c\) cuprates \cite{Emery}, 
are taken into account by the first term in the above hamiltonian,
and treated using the Monte Carlo technique.
Furthermore, this hamiltonian is easily developed so as to incorporate
an external magnetic field for the study of vortex states.   
Fourth, the superspin amplitude is fixed to unity in the above
hamiltonian.  The onset of superspin amplitude itself 
upon cooling can also be taken into 
account in the mean-field fashion, and is expected to
correspond to the so-called pseudo-gap phenomenon \cite{Zhang}.
Finally, only the simplest symmetry-breaking field \(g\) associated 
with the quadratic terms of order parameters is included in the hamiltonian.  Other symmetry-breaking fields appear when high orders of 
the order parameters are considered \cite{Zhang,Eder}.  
Although an argument on magnitudes of these fields is absent right
now, it is reasonably expected that the most important features of the 
breaking of \(SO(5)\) symmetry into the AF and SC subspaces are 
captured by the \(g\) field in (1).  

A typical simulation process starts from a random configuration of
superspins at a sufficiently high temperature.  The system is then
cooled gradually.  The equilibrium state at a given temperature is 
generated using typically 50,000 MC sweeps of update from the 
state of a slightly higher temperature.  In each sweep of
update, candidate vectors are generated randomly on the 
five-dimensional unit sphere for superspins on all sites in the system,
and are subject to the standard Metropolis algorithm to  
determine if they are accepted for the next configuration
\cite{Note1}.  After this equilibriation
process, statistics on physical quantities is performed over
100,000 MC sweeps.  Around transition temperatures,
more than \(10^6\) MC sweeps are spent in order to make sure
of sufficient equilibriation and statistics.  The system size for 
simulations in null external magnetic field is \(L^3=40^3\), 
with periodic boundary conditions in all crystal directions.
As the \(SO(5)\) superspins are continuous in five dimensions, and the 
system is of three dimensions in crystal space, a thorough analysis of 
finite-size effects 
on simulation results, which is important for determining the relevant 
critical and bicritical exponents in high precisions,
is extremely time consuming.   Only for several 
chosen parameter sets, larger systems have been simulated in order
to make sure that the main properties derived from the present
simulations do not suffer from finite-size effects. Systematic errors
(finite-size effects) are therefore not estimated for data 
presented in this paper.  Statistical errors are comparable to sizes
of marks in figures as far as not specified.  The AF coupling in the 
\(ab\) plane 
\(J^{AF}_{ab}\equiv J\) is taken as the energy unit, and temperature is 
measured by \(J/k_B\) throughout the present paper. 

\section{Phase diagrams and correlation functions for \(H=0\)}

\subsection{Isotropic system: 
\(J^{SC}_{ab}=J^{SC}_{c}=J^{AF}_c=J^{AF}_{ab}\equiv J\)}

Figure 1 is the temperature vs. \(g\)-field phase diagram of the 
system with the same AF and SC coupling in all
crystal directions.  Both the N/AF and
N/SC phase transitions are of second order, in the 3D Heisenberg 
and XY universality class, respectively.  The two phase
boundaries merge tangentially at the bicritical
point \([g_b,T_b]=[0, 0.85 J/k_B]\) \cite{Fisher}.  For \(g=g_b\)
and \(T>T_b\), the AF and SC correlation lengths for the two-point
correlation functions
are equal to each other, and isotropic in all crystal directions;
the weights of AF and SC components are  
\(3/5\) and \(2/5\), proportional to the number of degrees of freedom.
Away from the \(SO(5)\)-symmetric line, 
positive (negative) \(g\) fields suppress AF (SC) correlations at
all temperatures.  

\subsection{Anisotropic system: 
\(J^{SC}_{ab}=10J^{SC}_{c}=J^{AF}_c=J\)}

The temperature vs. \(g\)-field phase diagram of the system of couplings 
\(J^{SC}_{ab}=J^{AF}_{c}=J\) and \(J^{SC}_c=0.1J\) is presented in Fig. 2. 
The bicritical point is at \([g_b,T_b]=[1.18 J, 0.64 J/k_B]\).
The equal-weight partition of the superspin at \(g=g_b\)
observed in the isotropic system is broken.  Nevertheless, as indicated 
in the inset of Fig. 2, in the \(ab\) plane the AF correlation length is 
equal to the SC correlation length  when the \(g\) field is fixed at the
bicritical value.  This agreement is not trivial in contrast with the 
isotropic system.  The SC correlation length in the
 \(c\) axis is much smaller
than the other correlation lengths.  

\subsection{Strongly anisotropic system:
\(10J^{SC}_{ab}=100J^{SC}_{c}=100J^{AF}_{c}=J\)}

In order to simulate real high-\(T_c\) cuprates,  
the AF exchange coupling should be taken much stronger than the
effective SC coupling, and both AF and SC 
couplings are much weaker in the \(c\) axis.  
The temperature dependence of the AF staggered magnetization and 
the helicity modulus of the SC components 
\cite{Huc} are shown in Fig. 3 for the system of 
couplings \(J^{SC}_{ab}=0.1J\) and  \(J^{SC}_c=J^{AF}_c=0.01J\) at the
symmetry breaking field \(g=1.96 J\).  Since the helicity 
modulus is proportional to the superfluid density \cite{Fisher2},
 it is clear that the long-range SC order is established below the critical temperature 
\(T_c\simeq 0.115 J/k_B\).  As shown in the same figure,  the AF 
correlation length in the \(ab\) plane, \(\xi^{AF}_{ab}\),
increases at first as temperature is reduced, and then is suppressed 
as temperature approaches \(T_c\).  The maximal \(\xi^{SC}_{ab}\)
is taken at the temperature \(T_{sg}\simeq 0.15 J/k_B\).
The weight 
of AF components, \(\langle {\bf s}^2 \rangle\), decreases monotonically
in the whole cooling process and shows a sharp decline among
\(T_{sg}\) and \(T_c\).  Therefore, the enhancement of  \(\xi^{AF}_{ab}\) 
above \(T_{sg}\) is clearly the result of
reduction of thermal fluctuations; the suppression of 
\(\xi^{AF}_{ab}\) below \(T_{sg}\) is because of the loss of the AF
order in its competition with the SC order.  This peculiar 
behavior occurs because the AF coupling in the \(ab\) plane overwhelms 
over the SC one, while the SC 
groundstate is established by the large \(g\) field.  
Temperature dependence of the
 internal energy and the specific heat for this system
are depicted in Fig. 4.  No feature can be found around \(T_{sg}\)
in these two thermodynamic quantities.  Therefore, 
the only phase transition takes place at \(T_c\), and   
\(T_{sg}\) corresponds merely to a crossover. The SC correlation 
length in the \(ab\) plane, \(\xi^{SC}_{ab}\), diverges when temperature 
approaches \(T_c\) in Fig. 3, 
as usually in a thermodynamic second-order phase transition.

It is found experimentally that the spin-lattice relaxation
rate assumes its maximum at a temperature well above the SC
critical point \cite{Yasuoka}.  
The present simulation results indicate that this spin-gap
phenomenon can be explained by the competition among the long-range 
SC and AF orders, and thermal fluctuations.  
Since the enhancement of AF correlations above the SC critical
point is observed in the strongly anisotropic system of Fig. 3, but not in 
isotropic and slightly anisotropic systems of
Figs. 1, and 2, it becomes clear that in order to observe the spin-gap
behavior, the system should have SC couplings much weaker than
AF ones,  as in real high-\(T_c\) cuprates.

Figure 5 is the temperature vs. \(g\)-field phase 
diagram of the same couplings for Figs. 3 and 4.  
The bicritical point is at \([g_b,T_b]=[1.93 J, 0.12 J/k_B]\).
The latent heat associated with the first-order transition between the
AF and SC phases is approximately \(Q\simeq 0.05 J\), and decreases 
to zero as the bicritical point is approached.  The spin-gap
like phenomenon is observed in the region 
\(g_b < g < 2.2 J\).  For \(g>2.2 J\), AF correlations are suppressed by SC components at all temperatures.  
 The experimental fact that spin-gap behaviors are observed only in the 
underdoped region of high-\(T_c\) cuprates may be 
explained by the present simulation result.
The ratio between the spin-gap
temperature and the SC critical point is \(T_{sg}/T_c\simeq 1.6\)
at \(g=2 J\), which counts well the experimental 
observation \cite{Yasuoka}.   

In Fig. 5, the spin-gap temperature \(T_{sg}\) decreases 
as the bicritical point is approached.  This might seem curious 
at a first glance, since it is clear from hamiltonian (1)
that the larger the \(g\) field the smaller the AF components.
Shown in Figs. 6 (A) and (B) are the 
temperature dependence of the AF correlation length and the 
staggered susceptibility at several \(g\) fields.  Although both of them 
are monotonically 
suppressed by increasing \(g\) field when temperature is fixed, 
the temperature where they take maxima, \(T_{sg}\), increases with the
 \(g\) field, as clearly seen in Figs. 6.  It is noted that the spin-gap 
temperature increases with decreasing doping rate in experiments.  The 
present theory therefore conflicts with 
experimental observations in this aspect. 

The SC correlations are suppressed in the normal state above 
the AF phase boundary in the present system.  In this sense, there is no 
counterpart of the spin-gap temperature above N\'eel points.  
However, it is interesting to observe
in Fig. 7 that for the \(g\) field in a certain region below the bicritical
value, the SC weight, \(\langle {\bf t}^2\rangle\),
takes maximum at a temperature above the corresponding N\'eel point.
The temperature associated with the maximal SC weight, denoted by 
\(T_p\) in Fig. 5, may be identified with the pairing temperature 
\cite{Zhang}.  There is no feature in the internal
energy and the specific heat around this crossover temperature.

\section{Phase diagram and vortex states for \(H>H_{c1}\)}

\subsection{Model hamiltonian and phase diagram}

An external magnetic field penetrates into a type-II superconductor via 
thin flux lines associated with flux quanta for \(H\) larger than the 
lower critical field \(H_{c1}\).
SC is broken along the flux lines.  High-\(T_c\) superconductors are 
extremely type-II with very large Ginzburg-Landau numbers
\(\kappa\sim 100\).  
Research of the vortex states in high-\(T_c\) SC
has been growing into a vivid field of condensed matter physics 
and statistics.  The most important feature of the 
vortex states is that the Abrikosov FLL melts into FL 
liquid via a first-order phase transition \cite{Blatter,CN,Huc}. 

In the scheme of the \(SO(5)\) theory, the free energy of a vortex state 
can be reduced by rotating the superspins
from the SC subspace into the AF subspace at the flux-line cores 
\cite{Zhang}.   The possibility of AF cores of flux lines in the 
\(SO(5)\) theory was first addressed by Arovas {\it et al.} \cite{Arovas}. 
 Recently,  Alama {\it et al.} discussed
the \(\kappa\) dependence of the core state \cite{Alama}.   In 
these studies, the Abrikosov mean-field theory was developed so as to
incorporate the AF components.   
However, the Abrikosov mean-field theory for the vortex states is not
appropriate for the high-\(T_c\) SC since it only takes into account the 
amplitude of
SC order parameter, and cannot treat thermal fluctuations in the phase
variables, which are essentially important for determining the phase
diagram of the vortex states in high-\(T_c\) SC \cite{Huc}. 

The hamiltonian for the \(SO(5)\) theory in the presence
of an external magnetic field may be given as following \cite{Huso5}:

\begin{equation}
{\cal H}=
        -\sum_{\langle i,j\rangle} J^{SC}_{ij} |{\bf t}_i||{\bf t}_j|
           \cos\left(\varphi_i-\varphi_j-A_{ij}\right) 
        +\sum_{\langle i,j\rangle}J^{AF}_{ij}{\bf s}_i\cdot{\bf s}_j
        -\sum_i {\bf H}\cdot{\bf s}_i+g\sum_i {\bf s}^2_i,
   \quad A_{ij}=\frac{2\pi}{\phi_0}\int^{j}_{i}{\bf A}\cdot d{\bf r},
\label{eqn:H2}
\end{equation}

\noindent where \(|{\bf t}|\) and \(\varphi\) are the amplitude and phase 
of the SC order parameter.  The same constraint (2) is applied.
Fluctuations of the magnetic induction are neglected.  This
approximation is justified when the separation between vortices 
is larger than the SC correlation length
and much smaller than the penetration depth in the \(ab\) plane, a 
condition satisfied in large portion of \(H-T\) phase diagrams of
high-\(T_c\) superconductors.  The Josephson coupling should also be
 dominant over the electromagnetic coupling.  The first term in the above
hamiltonian, known as the fully frustrated 3D XY model, has been used
successfully for explaining many important thermodynamic properties of 
the vortex states in high-\(T_c\) SC \cite{Huc,Huab}.    

In order to simplify the situation, the case of an external magnetic field 
parallel to the \(c\) axis is addressed in the present paper \cite{Huab}.  
The vector potential is
given by \({\bf A}=(-yB/2,xB/2,0)\) with \(B=f\phi_0/l^2_{ab}\).
Here, \(\phi_0\) is the flux quantum, \(l_{ab}\) the unit
length in the \(ab\) plane, and \(f\) the average number of flux in 
each square unit cell in the \(ab\) plane. 
 The data shown in the following
are for \(f=1/25\), corresponding to the inter-vortex distance of
\(d_v=\sqrt{2/\sqrt{3}}l_{ab}/\sqrt{f}\simeq 5.37l_{ab}\) in the
triangular FLL.   The system size is chosen as 
\(L_a\times L_b\times L_c=50\times 50\times 40\) with periodic
boundary conditions in all crystal directions \cite{Huc}.  
Although the relation between the
magnetic induction \(B\) and the Zeeman field \(H\) is not clear, 
the value of the Zeeman field is not much relevant to the following 
discussions, and thus is fixed to \(H=0.1 J\).
In the present approach, vortices are defined by topological singularities
in the configuration of phase variables of SC order parameters: 
\(\sum_{\rm cell}(\varphi_i-\varphi_j-A_{ij})=(n-f)2\pi\), where \(n\)
is the vorticity.  

The temperature vs. \(g\)-field phase diagram of the 
system with couplings \(J^{SC}_{ab}=J^{AF}_c=J\) and 
\(J^{SC}_c=0.1J\) is depicted in Fig. 8.   
There are three ordered phases, namely the AF phase, AF and FLL 
coexistence phase, and FLL phase.
The onset of long-range SC order is a first-order phase transition,
same as those in systems of no AF components \cite{Huc}:  At the 
melting temperature \(T_m\) the FL liquid is frozen into the
triangular FLL; the helicity modulus along the \(c\)
axis jumps sharply from zero to a finite value; there is a 
\(\delta\)-function peak in the specific heat, associated with a 
small latent heat.  The onset of the long-range AF order is always
a second-order phase transition \cite{Note2}. 

From the comparison between the two phase diagrams in Figs. 2 
and 8 of same couplings,  it is clear that 
suppression of the transition temperature of the long-range SC order 
by the external magnetic field is much 
more significant than that of the long-range AF order. This difference is 
understood
easily considering the structures of these two long-range orders:
For the long-range AF order, the magnetic
spins are aligned antiferromagnetically, almost within the \(ab\) plane.  This configuration reduces the influence of the magnetic field on the 
onset of long-range AF order. On the other hand, the external magnetic 
field induces flux lines in the SC state, and produces
strong fluctuations in phase variables of SC order parameters.  
The long-range SC order is established
only when the flux lines are frozen into FLL.  The above difference in the
magnetic-field responses results also in the expansion of the AF phase 
into the SC territory, as can be seen in Figs. 2 and 8.

In the region \(1.0 J\le g\le 1.32 J\), the long-range AF and SC orders 
coexist at low temperatures.  The temperature dependence of the 
helicity modulus along the \(c\) axis, the staggered magnetization, and
the specific heat are shown in Fig. 9 for \(g=1.1 J\).  The N/AF 
transition at \(T_N\) is a thermodynamic second-order phase transition, 
above the first-order onset of the long-range SC order at \(T_m\).  
Although suppressed by the SC order in certain degree, the 
staggered magnetization survives to groundstate.

The AF phase boundary drops sharply from \([g, T]=[1.32 J, 0.59 J/k_B]\)
in Fig. 8.  The phase transition at this almost vertical part of AF phase
boundary is investigated by tuning the \(g\) field at a fixed temperature, 
in addition to the cooling process mentioned in Sec. II.  
The turning point on the N/AF phase boundary in Fig. 8
and the tricritical point in Ref.\cite{Huso5} are at the same 
temperature.

\subsection{AF vortex cores}

The vortex cores in the SC phase of the \(SO(5)\) theory are different 
from those without AF competition studied up to date.
The structure factors \(S({\bf q}_{ab}, z=0)\) for vortices,
\(s^2\), \(s_{ab}\), and \(s_c\) for \(g=1.5J\) in the FLL phase  are displayed in Figs. 10.   The Bragg peaks in the 
structure factor Fig. 10(A) for the vortex correlations 
are from the triangular FLL.  One also finds Bragg peaks in 
structure factor Fig. 10(B) for the AF amplitudes at the same 
wave numbers of Fig. 10(A).  This coincidence indicates clearly that
cores of flux lines are of larger AF weights than elsewhere.
The \(g\)-field dependence of the Bragg-spot height for AF weights in
the FLL phase, such as those in Fig. 10(B), is investigated when
temperature is fixed.  As shown in Fig. 11 for \(T=0.3J/k_B\),  
\(S({\bf q}_{ab}={\bf q}_{max}, z=0)\)
decays with increasing \(g\) field in a power law \(S\simeq p/g^q\)
with \(p=0.8\pm 0.05\) and \(q=3\pm 0.1\).  
The haloes at the wave numbers 
\(Q=(\pm \pi, \pm \pi)\) in the structure factors for \(s_{ab}\) and 
\(s_c\) in Figs. 10(C) and (D) correspond to short-range AF fluctuations. 
 The weak spot at \(Q=(0, 0)\)  in Fig. 10(D) is from the small 
ferromagnetic component \(s_c\) induced by the external magnetic 
field. 

The structure factors \(S({\bf q}_{ab}, z=0)\) in the AF and FLL 
coexistence phase 
for vortices, \(s^2\), \(s_{ab}\), and \(s_c\) are displayed  
in Figs. 12.  From the structure factors Figs. 12(A) and (B) for
vortices and AF amplitudes, it is clear 
that AF components are enhanced in cores of flux lines, as in the  
FLL phase.  In structure factors Figs. 12(C) and (D) for  \(s_{ab}\) and 
\(s_c\), there are strong Bragg peaks at \(Q=(\pm \pi, \pm \pi)\) 
associated with the long-range AF order.  
Satellite spots are observed around the 
main Bragg peaks in Figs. 12(C) and (D).  These satellite spots are easily
identified with those in Figs. 12(A) and (B).  Therefore, in the AF and FLL
coexistence phase, the phase of the 
long-range AF order is preserved in cores of flux lines.  

\section{Summary}

Thermodynamic properties of the \(SO(5)\) theory are 
investigated using Monte Carlo simulations on a model hamiltonian which
counts thermal fluctuations both in the rotations of superspins between
the SC and AF subspaces, and in the phase variables of the SC order 
parameters.  The latter factor is essentially important for explaining 
thermodynamic phase transitions associated with the onset of 
long-range SC order in high-\(T_c\) cuprates in null and finite
external magnetic fields.  Therefore, the present approach is superior to
Abrikosov-type mean-field treatments of the \(SO(5)\) theory, in which
thermal fluctuations in the phases of SC order parameters are neglected.

For null external magnetic field, there is a
bicritical point in the temperature vs. \(g\)-field phase diagram, at
which the second-order N/AF and N/SC phase boundaries merge 
tangentially into the first-order AF/SC phase boundary.
Hysteresis phenomenon is observed at the
first-order AF/SC phase transition, which may suggest
a phase-separation region in the phase diagram.  

In systems with much stronger AF couplings than SC ones while 
the SC groundstate is achieved by \(g\) fields larger than the bicritical
value,  AF correlations are enhanced at a crossover temperature above the 
SC critical point.  The origin of this enhancement in AF 
correlations is clarified to be the competition among the long-range
AF and SC orders and thermal fluctuations.  When the \(g\) field becomes
too large, this crossover fades away since AF correlations  are
suppressed at all temperatures by the large SC component.
These results are consistent with the spin-gap 
phenomenon observed experimentally in the following aspects:  
First, real cuprates are very anisotropic in the AF
and SC couplings \(J^{AF}\sim 0.1\) eV and \(J^{SC}\sim 0.01\) eV;  
Second, the spin-gap phenomenon
has been observed experimentally only in the underdoped region. 
In contrast with experimental observations, however, the spin-gap 
temperature decreases as the bicritical point is approached from the SC
side.  Near the bicritical point, there is a crossover temperature above 
the  N\'eel temperature where the weight of SC components takes 
maximum.

The \(SO(5)\) theory in an external magnetic field is also investigated.
The long-range SC order is established 
through a first-order freezing transition from the FL liquid into the FLL,
while the onset of the long-range AF order is associated with a 
second-order phase transition.  These two phase boundaries cross
each other, and thus produce a region in the phase diagram where the two 
long-range orders coexist.  
In the FLL phase, only short-range AF fluctuations
are enhanced at cores of flux lines. 1D long-range AF order along the flux 
line cannot be realized because of strong thermal fluctuations. 
In the coexistence phase, superlattice spots surrounding the strong AF 
Bragg peaks at \(Q=(\pm\pi,\pm\pi)\) are observed in the simulated structure factor, and are identified with the modulation by the 
triangular flux-line lattice of SC.  This simulation result can be 
checked by the neutron scattering technique.   

\vskip2cm
 
\begin{center}
{\bf Acknowledgements}
\end{center}

The author would like to thank S.-C. Zhang, T. Koyama,  and Y. K. Bang for 
stimulating conversations on the \(SO(5)\) theory.  
M. Tachiki is very grateful for drawing author's attention to vortex 
states in high-\(T_c\) superconductivity and continuous encouragement. 
He appreciates S. Miyashita, N. Akaiwa, M. Itakura, and Y. Nonomura for 
helpful discussions on technical points of MC simulation.
The present simulations are performed on the Numerical Materials 
Simulator (SX-4) of National Research Institute for Metals (NRIM), Japan.

\newpage

%\vskip3cm

\noindent Figure Captions

\noindent Fig. 1: Temperature vs. \(g\)-field phase diagram of the 
system with isotropic AF and SC couplings.  
The bicritical point is at \([g_b,T_b]=[0, 0.85J/k_B]\). 

\noindent Fig. 2: Temperature vs. \(g\)-field phase diagram of the 
system with the couplings \(J^{SC}_{ab}=J^{AF}_c=J\), and 
\(J^{SC}_c=0.1J\).  The bicritical point is at 
\([g_b,T_b]=[1.18 J, 0.64J/k_B]\).  Inset: temperature 
dependence of the AF and SC correlation lengths in the \(ab\) plane
at \(g=g_b\).  

\noindent Fig. 3: Temperature dependence of the AF and SC 
order parameters, correlation lengths, and the weight of AF components
at \(g=1.96J\) in the system with couplings \(J^{SC}_{ab}=0.1J\) and
\(J^{SC}_c=J^{AF}_c=0.01J\).  Here, \(T_c\) is the SC transition point,
and \(T_{sg}\) is the spin-gap temperature.

\noindent Fig. 4: Temperature dependence of the internal energy and the
specific heat per site for the same system in Fig. 3.

\noindent Fig. 5: Temperature vs. \(g\)-field phase diagram for the 
same couplings in Fig. 3.  
The bicritical point is at \([g_b,T_b]=[1.93J, 0.12J/k_B]\).
The spin-gap temperature \(T_{sg}\) and pairing temperature \(T_p\)
fade away around \(g=2.2J\) and \(g=1.5J\) respectively.

\noindent Fig. 6: Temperature dependence of the AF correlation 
length in the \(ab\) plane (A) and the staggered susceptibility (B) at
several typical \(g\) fields.  Maxima are assumed
at the spin-gap temperatures \(T_{sg}\) for \(g>g_b=1.93J\).

\noindent Fig. 7: Temperature dependence of the SC weight at a series of
\(g\) fields.  Maxima are assumed at the pairing
temperatures \(T_p\) for \(1.5J<g<g_b\).

\noindent Fig. 8:  Temperature vs. \(g\)-field phase diagram of the 
system with the couplings \(J^{SC}_{ab}=J^{AF}_c=J\) and 
\(J^{SC}_c=0.1J\). The flux density is given by
 \(f=1/25\), and the Zeeman field is \(H=0.1J\).

\noindent Fig. 9:  Temperature dependence of the helicity modulus along
the \(c\) axis, the staggered magnetization, and the specific heat per
site in the system of the same couplings of Fig. 8 at \(g=1.1J\).

\noindent Fig. 10: Structure factors \(S({\bf q}_{ab}, z=0)\) for vortices (A), \(s^2\) (B), \(s_{ab}\) (C), and \(s_c\) (D) in the FLL phase in Fig. 8.

\noindent Fig. 11: \(g\)-field dependence of the Bragg-spot height for the 
AF weights at \(T=0.3J/k_B\). 

\noindent Fig. 12: Structure factors \(S({\bf q}_{ab}, z=0)\) for vortices
(A), \(s^2\) (B), \(s_{ab}\) (C), and \(s_c\) (D) in the AF and FLL 
coexistence phase in Fig. 8.

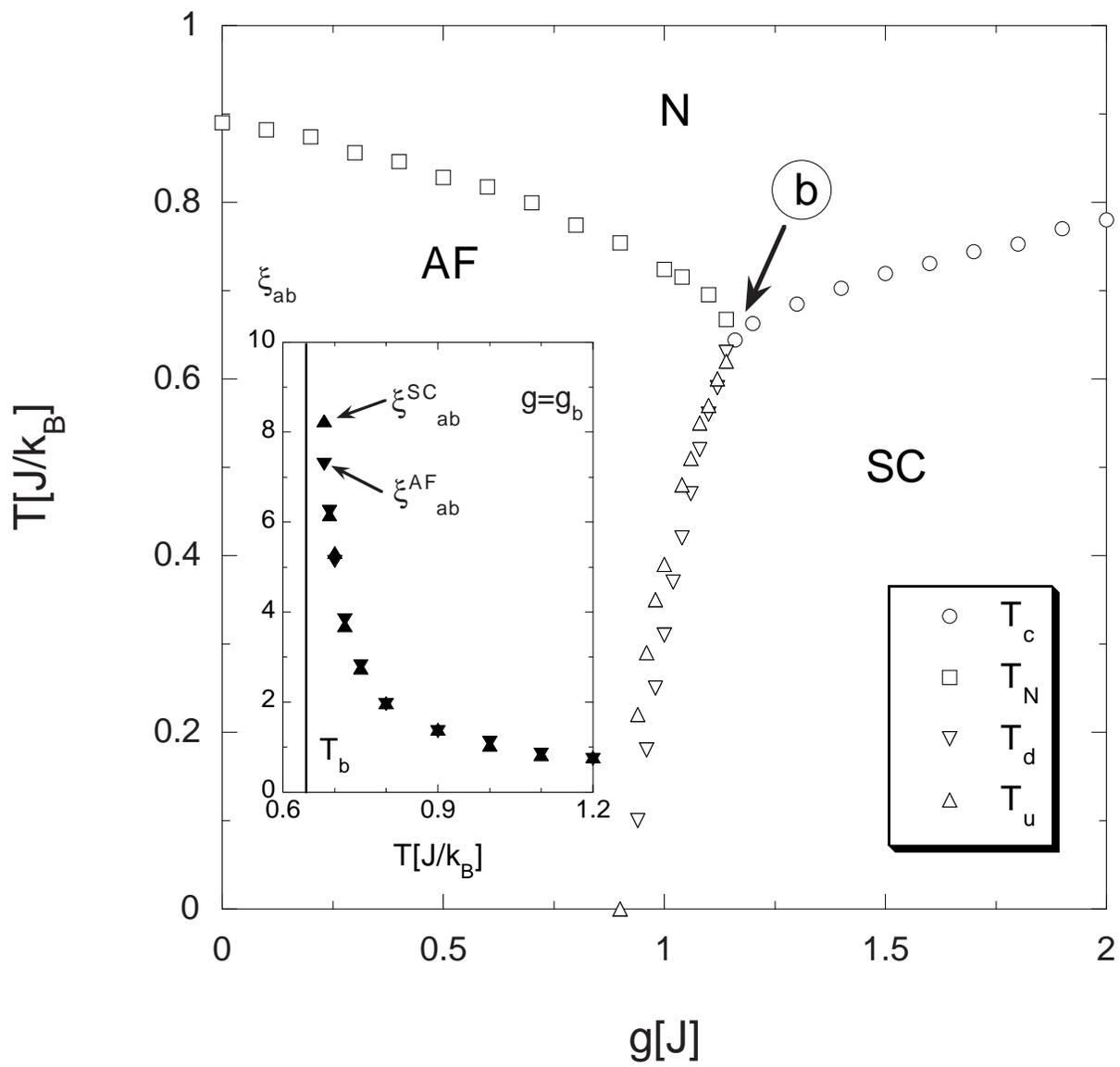
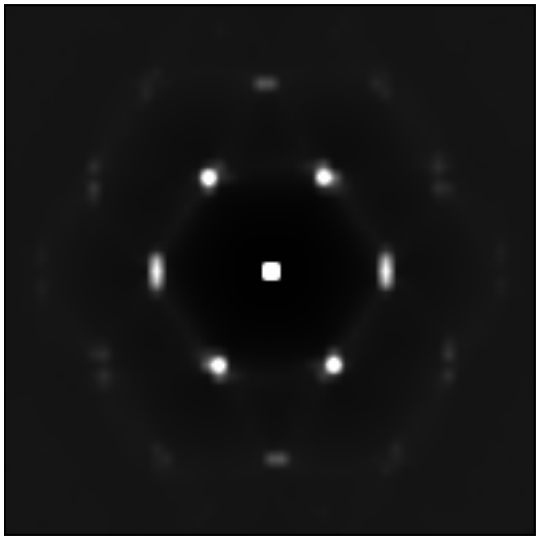
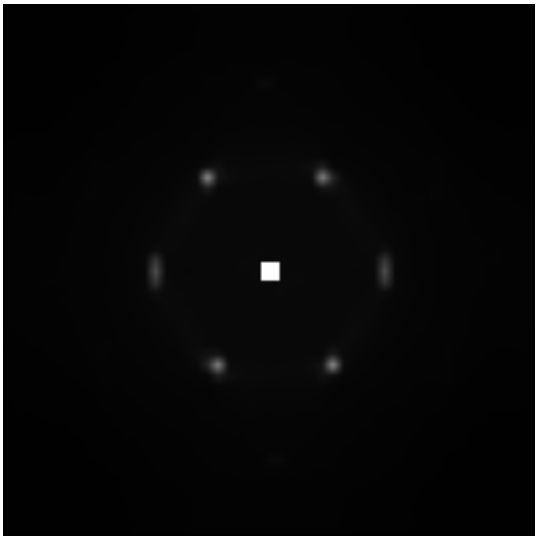
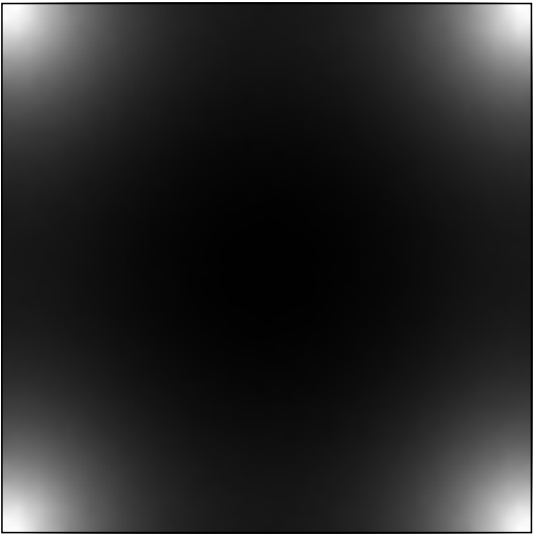
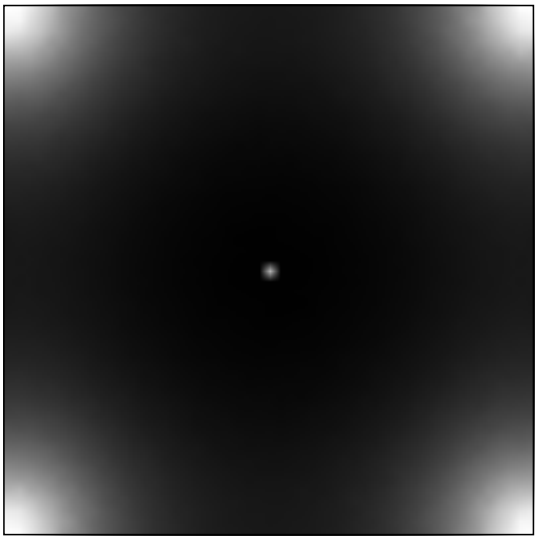
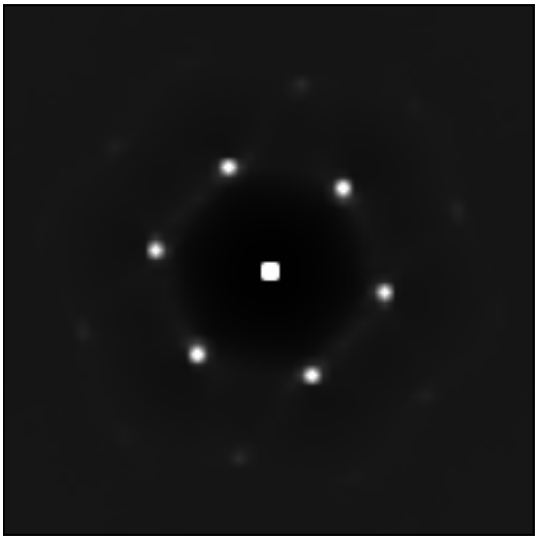
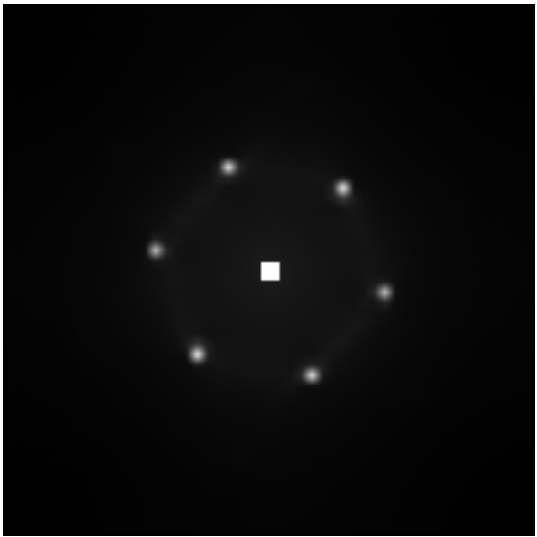
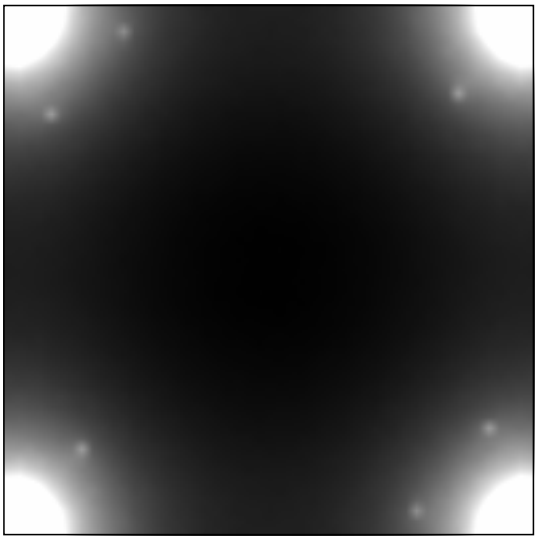
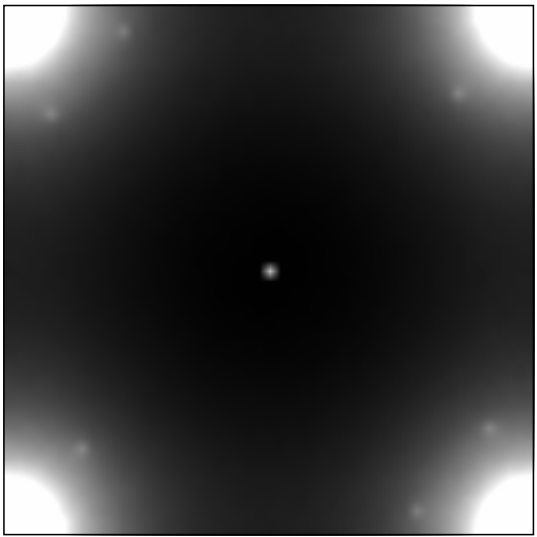
\end{document}